\documentclass[epj]{svjour}
\usepackage{graphicx}

\begin{document}

\title{Minimal conductivity in bilayer graphene}
\titlerunning{Minimal conductivity in bilayer graphene}
\author{M. I. Katsnelson}
\institute{Institute for Molecules and Materials, Radboud
University Nijmegen, 6525 ED Nijmegen, The Netherlands}
\date{Received: date / Revised version: date}

\abstract {Using the Landauer formula approach, it is proven that
minimal conductivity of order of $e^{2}/h$ found experimentally in
bilayer graphene is its intrinsic property. For the case of ideal
crystals, the conductivity turns our to be equal to $e^{2}/2h$ per
valley per spin. A zero-temperature shot noise in bilayer graphene
is considered and the Fano factor is calculated. Its value
$1-2/\pi$ is close to the value 1/3 found earlier for the
single-layer graphene.
     }

\maketitle

It has been observed recently that bilayer graphene, that is a
two-dimensional allotrope of carbon  formed by two graphite atomic
sheets, has a minimal conductivity of order of $e^{2}/h$
\cite{natphys}. The same property has been found earlier in the
single-layer graphene \cite {kostya2,kim}. Both single- and
bilayer graphene are gapless semiconductors, with conical and
parabolic touching of electron and hole bands, respectively
\cite{natphys,kostya2,kim}. The charge carriers in the
single-layer graphene are massless Dirac fermions which is a
crucial point when explaining the conductivity minimum
\cite{ktsn,been,zieg}. Actually, this anomalous property of the
two-dimensional massless fermions was considered theoretically
\cite{D1,D2,ludwig} before discovery of graphene. A crucial
physical phenomenon here is the Zitterbewegung of quantum
ultrarelativistic particles \cite{ktsn} which plays a role of
``intrinsic'' disorder; the latter being confirmed by calculations
of the shot noise in \textit{ideal} graphene for zero doping which
turns out to have the same value (Fano factor 1/3) as
\textit{disordered} metals \cite{been}. At the same time,
observation of the finite minimal conductivity in bilayer graphene
is a serious challenge for theory \cite{natphys}. Here I present a
solution of this problem based on the same Landauer formula
approach which was used earlier for the single-layer case
\cite{ktsn,been}.

The bilayer graphene is a zero-gap semiconductor with \textit{parabolic}
touching of the electron and hole bands described by the single-particle
Hamiltonian \cite{natphys,falko}
\begin{equation}
H=\left(
\begin{array}{cc}
0 & -\left( p_{x}-ip_{y}\right) ^{2}/2m \\
-\left( p_{x}+ip_{y}\right) ^{2}/2m & 0
\end{array}
\right)   \label{bilayer}
\end{equation}
where $p_{i}=-i\hbar \partial /\partial x_{i}$ are electron
momenta operators and $m$ is the effective mass  (here we ignore
some complications due to large-scale hopping processes which are
important for a very narrow range of the Fermi energies
\cite{falko}). Two components of the wave function are originated
from crystallographic structure of graphite sheets with two carbon
atoms in the sheet per elementary cell. There are two touching
points per Brillouin zone, $K$ and $K^{\prime }$. For ideal
crystals, no Umklapp processes between these points are allowed
and thus they can be considered independently. Our final result
for the conductivity should be just multiplied by four due to two
touching points and two spin projections (we will not take into
account electron spin explicitly in our consideration). To
calculate the conductivity at zero energy we will use the Landauer
formula expressing the conductance of the system in terms of
transmission coefficients. Similar to Ref. \cite{ktsn} we will use
the simplest boundary conditions assuming that the sample is a
ring of length $L_{y}$ in the $y$-direction  and leads connected
with the sample at $x=0$ and $x=L_{x}$ are made from doped bilayer
graphene with potential $V_{0}>0$ and Fermi energy
$E_{F}=-V_{0}=-\hbar ^{2}k_{F}^{2}/2m$ (Figure 1).
\begin{figure}[tbp]
\includegraphics[width=9cm]{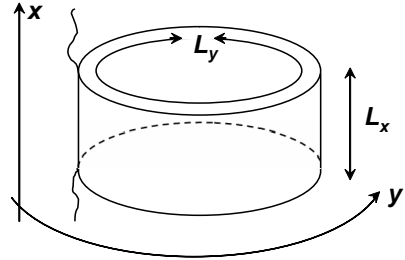}
\caption{Geometry of the sample.} \label{fig:1}
\end{figure}

Let us first find the solution of the Schr\"{o}dinger equation
with zero energy,
$H\Psi =0$ where $\Psi $ is a ``spinor'' with components $\psi _{1}$ and $%
\psi _{2}$. They satisfy the equations
\begin{eqnarray}
\left( \frac{\partial }{\partial x}-i\frac{\partial }{\partial y}\right)
^{2}\psi _{2} &=&0,  \label{psi2} \\
\left( \frac{\partial }{\partial x}+i\frac{\partial }{\partial y}\right)
^{2}\psi _{1} &=&0.  \label{psi1}
\end{eqnarray}
Due to the periodicity in the $y$ direction both wave functions
should be proportional to $\exp \left( ik_{y}y\right) $ where
$k_{y}=2\pi
n/L_{y},n=0,\pm 1,\pm 2,....$ This immediately gives us the following $x$%
-dependence for the wave functions:
\begin{eqnarray}
\psi _{1}\left( x\right)  &=&\left( A_{1}x+B_{1}\right) e^{k_{y}x},
\nonumber \\
\psi _{2}\left( x\right)  &=&\left( A_{2}x+B_{2}\right) e^{-k_{y}x}
\label{sol1}
\end{eqnarray}
$(0<x<L_{x}).$ The constants $A_{i}$ and $B_{i}$ should be found
from the boundary conditions at $x=0$ and $x=L_{x}$ which are
nothing but continuity conditions for both functions $\psi _{1}$
and $\psi _{2}$ and their derivatives \cite{klein}.

It will be shown further that the values of $k_{y}$ essential for
the electron transmission are of the order of $L_{y}^{-1}$ and
thus much smaller than $k_{F}$ in the leads. Therefore, one can
restrict ourselves to the case of normal incidence only for the
wave functions outside the sample:
\begin{eqnarray}
\psi _{1}\left( x\right)  &=&e^{ik_{F}x}+re^{-ik_{F}x}+ce^{k_{F}x},
\nonumber \\
\psi _{2}\left( x\right)  &=&e^{ik_{F}x}+re^{-ik_{F}x}-ce^{k_{F}x}
\label{sol2}
\end{eqnarray}
for $x<0$ and

\begin{eqnarray}
\psi _{1}\left( x\right)
&=&te^{ik_{F}(x-L_{x})}+de^{-k_{F}(x-L_{x})},
\nonumber \\
\psi _{2}\left( x\right)
&=&te^{ik_{F}(x-L_{x})}-de^{-k_{F}(x-L_{x})} \label{sol3}
\end{eqnarray}
for $x>L_{x}.$ Here $r$ and $t$ are reflection and transmission
coefficients, respectively. One should stress that to satisfy all
the boundary conditions for the case of a bilayer, one has to
include not only oscillatory but also exponentially decaying
solutions of the Schr\"{o}dinger equation \cite{klein}.

Using the boundary conditions at the sample-lead boundary, one
finds the set of linear equations
\begin{eqnarray}
1+r+c &=&B_{1},  \nonumber \\
1+r-c &=&B_{2},  \nonumber \\
k_{F}\left[ i\left( 1-r\right) +c\right]  &=&A_{1}+B_{1}k_{y},  \nonumber \\
k_{F}\left[ i\left( 1-r\right) -c\right]  &=&A_{2}-B_{2}k_{y},  \nonumber \\
F_{1}X &=&t+d,  \nonumber \\
F_{2}X^{-1} &=&t-d,  \nonumber \\
\left( F_{1}k_{y}+A_{1}\right) X &=&k\left( it-d\right) ,  \nonumber \\
\left( -F_{2}k_{y}+A_{2}\right) X^{-1} &=&k\left( it+d\right)
\label{system}
\end{eqnarray}
where $F_{i}=A_{i}L_{x}+B_{i},X=\exp \left( k_{y}L_{x}\right) .$

By use of the assumptions
\begin{equation}
k_{F}\gg k_{y},L_{x}^{-1}  \label{assum}
\end{equation}
one can easily solve the equations (\ref{system}) and find
\begin{equation}
t=\frac{2iL_{x}}{k_{F}}\frac{\cosh \left( k_{y}L_{x}\right) }{L_{x}^{2}+%
\frac{2i}{k_{F}^{2}}\cosh ^{2}\left( k_{y}L_{x}\right) }.  \label{t}
\end{equation}
Corrections to this formula are of order of $1/(k_F L_x)$; we
cannot keep them in the answer since terms of the same order of
magnitude have been omitted by considering the normal-incidence
case for the wave functions in the leads (\ref{sol2}),
(\ref{sol3}).

Thus, for the transmission coefficient $T_{n}=\left| t\left(
k_{y}=2\pi n/L_{y}\right) \right| ^{2}$ one obtains the final
result
\begin{equation}
T_{n}=\frac{4k_{F}^{2}L_{x}^{2}\cosh ^{2}\left( k_{y}L_{x}\right) }{%
k_{F}^{4}L_{x}^{4}+4\cosh ^{4}\left( k_{y}L_{x}\right) }.  \label{Tn}
\end{equation}

One can see that the transmission coefficient reaches the maximum
value equal to 1 at $\cosh \left( k_{y}L_{x}\right)
=k_{F}L_{x}/\sqrt{2}$ or, approximately, at
\begin{equation}
k_{y}/k_{F}\simeq \ln \left( \sqrt{2}k_{F}L_{x}\right) /\left(
k_{F}L_{x}\right)   \label{angle}
\end{equation}
which obviously satisfies the condition (\ref{assum}) for
macroscopically large $L_{x}\gg k_{F}^{-1}.$ Note that the
complete transmission through the potential barrier for some
finite incident angles is a characteristic property of the bilayer
case, in contrast with the single layer, where the complete
transmission takes place at exactly normal incidence \cite{klein}.

Using the Landauer formula (for review, see Refs.
\cite{beenakker,buttiker}) one can calculate the conductance per
valley per spin
\begin{equation}
g=\frac{e^{2}}{h}\sum\limits_{n=-\infty }^{\infty }T_{n}.  \label{land}
\end{equation}
Similar to Refs. \cite{ktsn,been} to calculate the conductivity of
bilayer graphene at zero energy one should consider the case
$L_{y}\gg L_{x}.$ In that case the sum in Eq.(\ref{land}) can be
replaced by an integral. Introducing the integration variable
$z=\cosh \left( 2 k_{y}L_{x}\right) +1 $ and taking into account
the condition (\ref{assum}) one finds for the conductivity $\sigma
=\left( L_{x}/L_{y}\right) g$:
\begin{equation}
\sigma =\frac{e^{2}}{2h}.  \label{answer}
\end{equation}
Thus, the conductivity of bilayer graphene has the same order of
magnitude than for the single-layer case (where the coefficient
$1/\pi $, instead of 1/2, was obtained by similar method in Refs.
\cite{ktsn,been}). This result looks rather unexpected since the
electron spectra in these two cases are drastically different.
More accurate calculations of the integral gives a correcting
multiplier $1+ \frac{4 \ln \left(k_F L_x \right)}{\pi k_{F}^{2}
L_{x}^{2}}+ ...$ in Eq.(\ref{answer}).

Following Ref. \cite{been} one can estimate the Fano factor
characterizing the intensity of electron shot noise:
\begin{equation}
F=\frac{\sum\limits_{n=-\infty }^{\infty }T_{n}\left( 1-T_{n}\right) }{%
\sum\limits_{n=-\infty }^{\infty }T_{n}}  \label{fano}
\end{equation}
(for a general review of the quantum-limited shot noise and
physical meaning of the Fano factor, see Refs.
\cite{buttiker,phystoday}). A straightforward calculation for the
case  $L_{y}\gg L_{x}$ gives us the answer
\begin{equation}
F=1-\frac{2}{\pi }  \label{f1}
\end{equation}
which is rather close to the value 1/3 found for the case of the
single layer, as well as for the case of disordered metals
\cite{been}. This means that, in a sense, the case of bilayer
graphene is also characterized by some ``intrinsic'' disorder
similar to the Zitterbewegung \cite{ktsn}.

Unfortunately, the accuracy of experimental data \cite{natphys} is
not sufficient to establish the numerical coefficient in the
expression for the minimal conductivity. For the case of a single
layer it is close to 1, instead of $1/\pi$.

To conclude, we have demonstrated that Dirac energy spectrum is
actually not important for existence of minimal conductivity in
graphene. The latter has the same order of magnitude both for
conical (a single-layer case) and for parabolic (a bilayer case)
energy spectrum near the band crossing point.

I am thankful to Andre Geim and Kostya Novoselov for valuable
discussions stimulating this work. This work was supported by the
Stichting voor Fundamenteel Onderzoek der Materie (FOM) which is
financially supported by the Nederlandse Organisatie voor
Wetenschappelijk Onderzoek (NWO).

\end{document}